\documentstyle[amssymb,12pt]{article}
\begin{document}

\begin{center}
{\bf Reliability Function of General Classical-Quantum Channel} \vskip20pt {
A.~S.~Holevo} \footnote{
Visiting Professor at the Institute for Mathematical Physics, Technical
University of Braunschweig, permanently with Steklov Mathematical Institute,
Russian Academy of Sciences.} \footnote{
Work supported by A. von Humboldt Foundation, Isaac Newton Institute and ESF.
} \footnote{
The paper was partially reported at the Workshop on Complexity, Computation
and the Physics of Information, Cambridge, July 1999.}
\end{center}

\vskip30pt {\small {\bf Abstract -- In information theory the reliability
function and its bounds, describing the exponential behavior of the error
probability, are important quantitative characteristics of the channel
performance. From a more general point of view, these bounds provide certain
measures of distinguishability of a given set of classical states. In the
paper \cite{bur} quantum analogs of the random coding and the expurgation
lower bounds for the case of pure signal states were introduced. Here we
discuss the case of general quantum states, in particular, we prove
the expurgation bound conjectured in \cite{bur} and find the quantum cutoff
rate for arbitrary mixed signal states. \vskip10pt Index Terms -- Quantum
channel, reliability function, random coding, expurgation.}} \newpage

\begin{center}
\centerline{\sc I. Introduction}
\end{center}

We consider classical-quantum channel \cite{hol73} with a finite input
alphabet $\{1,...,a\}$ and with arbitrary signal states given by density
operators $S_{i};\ i=1,...,a$ in a Hilbert space ${\cal H}$. For simplicity
of presentation we take ${\cal H}$ finite-dimensional, although with obvious
modifications the results hold for a separable ${\cal H}$. The classical
channel corresponds to the case of commuting operators $S_{i}$, represented
by diagonal matrices $diag\left( P(1|i),\dots
,P(b|i)\right) ,$ where  $P(j|i)$ is the channel transition probability.

Product channel of degree $n$ acts in the tensor product ${\cal H}
^{\otimes n}={\cal H}\otimes \ldots \otimes {\cal H}$ of $n$ copies of the
space ${\cal H}$. Sending a codeword $w=(i_{1},\ldots
,i_{n}),i_{k}\in \{1,\ldots a\},$ produces the signal state $
S_{w}=S_{i_{1}}\otimes \ldots \otimes S_{i_{n}}$ in the space ${\cal H}
^{\otimes n}$. A code $({\cal W},{\bf X})$ of size $M$ in ${\cal
H}^{\otimes n}$ is a collection of $M$ pairs $(w^{1},X_{1}),\ldots
,(w^{M},X_{M}),$ where ${\cal W=}\left\{ w^{1},\ldots ,w^{M}\right\} $ is
a codebook, ${\bf X}=\{X_{1},\ldots ,X_{M}\}$ is a quantum decision
rule, i.e. a collection of positive operators in ${\cal H}^{\otimes n},$
satisfying $\sum_{j=1}^{M}X_{j}\leq I$ \cite{hol73}. The conditional
probability to make a decision in favor of message $w^{k}$ provided
that codeword $w^{j}$ was transmitted is ${\rm Tr}S_{w^{j}}X_{k},$ in
particular, the probability to make wrong decision is equal to
\[
P_{j}({\cal W},{\bf X})=1-{\rm Tr}S_{w^{j}}X_{j}.
\]

One usually considers the error probabilities
\[
P_{max}({\cal W},{\bf X})=\max_{1\leq j\leq M}P_{j}({\cal W},{\bf X})
\]
and
\[
\bar{P}({\cal W},{\bf X})={\frac{1}{M}}\sum_{j=1}^{M} P_{j}({\cal W},{\bf X}).
\]
We shall denote by $P_{e}(M,n)$ any of the minimal error probabilities
\newline
$\min_{{\cal W},{\bf X}}P_{max}({\cal W},{\bf X}),\min_{{\cal W},{\bf X}}
\bar{P}({\cal W},{\bf X}).$ It is known that they are essentially equivalent
from the point of view of information theory \cite{gal}, see also Sect. 3.

The {\sl classical capacity} of the classical-quantum channel is defined as
the number $C$ such that $P_{e}(\mbox{2}^{nR},n)$ tends to zero as $
n\rightarrow \infty $ for any $0\leq R<C$ and does not tend to zero if $
R>C$. Moreover, if $R<C$ then $P_{e}(\mbox{2}^{nR},n)$ tends to zero
exponentially with $n$ and we are interested in the logarithmic rate of
convergence given by the{\sl \ reliability function}
\begin{equation}
E(R)=-\lim_{n\rightarrow \infty }\inf {\frac{1}{n}}\log P_{e}(2
^{nR},n),\quad 0<R<C.  \label{f1}
\end{equation}
In the classical information theory \cite{gal} there are lower and upper
bounds for $E(R)$, giving important quantitative characteristics of the
channel performance. From a more general point of view, these bounds provide
certain measures of distinguishability of a given set of classical states.
In the paper \cite{bur} quantum analogs of the random coding and the
expurgation lower bounds were given for the case of {\sl pure} signal states
$S_{i},$ represented by rank one density operators. Here we discuss the
general case, in particular, we prove the expurgation bound conjectured in
\cite{bur}.

\begin{center}
\centerline{\sc II. The capacity and the random coding lower bound}
\end{center}

The classical capacity of the channel is given by the formula
\begin{equation}
C=\max_{\pi }\left[ H\left( \sum_{i=1}^{a}\pi _{i}{S}_{i}\right)
-\sum_{i=1}^{a}\pi _{i}H({S}_{i})\right] ,  \label{cap}
\end{equation}
where $H(S)=-\mbox{Tr}S\log S$ is the von Neumann entropy of the state $S$
and $\pi =\{\pi _{i}\}$ are probability distributions on the input alphabet $
\{1,...,a\}$. This relation was established in \cite{hol96}, \cite{1schum},
using the concept of typical subspace \cite{schum}. The proofs in the present paper
are direct, making no use of this concept and of the relation (\ref{cap}).

{\sl Proposition 1:} For any $\pi $ and $0<s\leq 1$
\begin{equation}
H\left( \sum_{i=1}^{a}\pi _{i}S_{i}\right) -\sum_{i=1}^{a}\pi
_{i}H(S_{i})\geq \frac{1}{s}\mu (\pi ,s),  \label{cap1}
\end{equation}
where
\[
\mu (\pi ,s)=-\log \mbox{Tr}\left( \sum_{i=1}^{a}\pi _{i}S_{i}^{\frac{1}{1+s}
}\right) ^{1+s}.
\]

{\sl Proof.} Denote by
\[
H(S,T)=\left\{
\begin{array}{cc}
\mbox{Tr}S(\log S-\log T), & \mbox{ if supp}S\subseteq \mbox{ supp}T, \\
+\infty  & \mbox{otherwise}
\end{array}
\right.
\]
\[
H_{r}(S,T)=-\log \mbox{Tr}S^{1-r}T^{r};\quad 0\leq r\leq 1,
\]
the relative entropy and the Chernoff-R\'{e}nyi entropy of the density
operators $S,T$, correspondingly (see \cite{petz}). Since $H(S,T)=\left.
\frac{d}{dr}\right| _{r=0}H_{r}(S,T),$ and $H_{r}(S,T)$ is concave, we have $
rH(S,T)\geq H_{r}(S,T).$ Now
\[
H\left( \sum_{i=1}^{a}\pi _{i}S_{i}\right) -\sum_{i=1}^{a}\pi
_{i}H(S_{i})=\sum_{i=1}^{a}\pi _{i}H(S_{i},\sum_{l=1}^{a}\pi _{l}S_{l})
\]
\[
\geq -\frac{1}{r}\sum_{i=1}^{a}\pi _{i}\log \mbox{Tr}S_{i}^{1-r}\left(
\sum_{l=1}^{a}\pi _{l}S_{l}\right) ^{r}\geq -\frac{1}{r}\log \mbox{Tr}
\sum_{i=1}^{a}\pi _{i}S_{i}^{1-r}\left( \sum_{l=1}^{a}\pi _{l}S_{l}\right)
^{r},
\]
by convexity of $-\log .$ By quantum H\"{o}lder inequality \cite{scha}, the argument of $
\log $ is less than or equal to
\[
\left( \mbox{Tr}\left( \sum_{i=1}^{a}\pi _{i}S_{i}^{1-r}\right) ^{p}\right)
^{\frac{1}{p}}\left( \mbox{Tr}\left( \sum_{l=1}^{a}\pi _{l}S_{l}\right)
^{rq}\right) ^{\frac{1}{q}}
\]
if $p^{-1}+q^{-1}=1,p>1.$ Putting $p=\frac{1}{1-r},q=\frac{1}{r},s=\frac{r}{
1-r},$ and using monotonicity of $\log ,$ we obtain (\ref{cap1}).
$\Box $

Assume now that the words in the codebook ${\cal W}$ are chosen at random,
independently, and with the probability distribution
\begin{equation}
{\sf P}\{w=(i_{1},...,i_{n})\}=\pi _{i_{1}}\cdot ...\cdot \pi _{i_{n}}
\label{iii5}
\end{equation}
for each word. We shall denote expectations with respect to this probability
distribution by the symbol ${\sf E.}$ In \cite{bur} we conjectured the
following {\sl random coding bound} for the error probability
\begin{equation}
{\sf E}\min_{{\bf X}}\bar{P}({\cal W},{\bf X})\leq c\inf_{0<s\leq 1}(M-1)^{s}
\left[ \mbox{Tr}\left( \sum_{i=1}^{a}\pi _{i}S_{i}^{\frac{1}{1+s}}\right)
^{1+s}\right] ^{n}.  \label{f27}
\end{equation}
The bound (\ref{f27}) holds for pure states $S_{i}$ in which case $S_{i}^{
\frac{1}{1+s}}=S_{i}$ and $c=2$ \cite{bur}. For commuting $S_{i}$ it reduces
to the classical bound of Theorem 5.6.2 \cite{gal} with $c=1$. By putting $
M=2^{nR},$ it implies the lower bound for the reliability function
\begin{equation}
E(R)\geq \max_{\pi }\sup_{0<s\leq 1}[\mu (\pi ,s)-sR]\equiv E_{r}(R).
\label{er1}
\end{equation}
This can be calculated explicitly for quantum binary and Gaussian pure state
channels \cite{hol73}. A remarkable feature of the classical case is that
there exists the upper bound (the sphere-packing bound) which coincides with
$E_{r}(R)$ for high rates, and thus gives exact expression for $E(R).$ In
the quantum case no useful upper bound for $E(R)$ is known yet (see,
however, \cite{wint} for an incomplete analog of the sphere-packing bound).

We shall prove a general inequality for the error probabilities $P_{j}({\cal
W},{\bf X}),$ which implies (\ref{f27}) for $s=1$ with $c=1,$ and will be
used in the next Section to obtain the expurgation bound. Moreover, the
first part of the argument will be used for alternative operator proof of (
\ref{f27}) in case of commuting ${S}_{i},$ indicating clearly at which point
commutativity comes into play. The proof of (\ref{f27}) in full generality
remains open question.

{\sl Lemma:} For any collection ${\cal W}$ ${\cal \ }$ of codewords there is
a decision rule ${\bf X}$ such that
\begin{equation}
P_{j}({\cal W},{\bf X})\leq \mbox{Tr}\sqrt{S_{w^{j}}}\sum_{l\not=j}\sqrt{
S_{w^{l}}},\quad j=1,\dots M.  \label{e5}
\end{equation}

{\sl Proof.} By making a small perturbation of the density operators $
S_{w^{j}}$, we can assume that they are nondegenerate. We choose the
following suboptimal decision rule
\begin{equation}
X_{j}=(\sum_{l=1}^{M}S_{w^{l}}^{\quad r})^{-1/2}S_{w^{j}}^{\quad
r}(\sum_{l=1}^{M}S_{w^{l}}^{\quad r})^{-1/2},  \label{gs1}
\end{equation}
where $r$ is a real parameter, $0<r\leq 1$. This gives
\begin{equation}
P_{j}({\cal W},{\bf X})=1-\mbox{Tr}S_{w^{j}}{A_{j}}^{\ast }{A_{j}},
\end{equation}
where ${A_{j}}=S_{w^{j}}^{\quad r/2}(\sum_{l=1}^{M}S_{w^{l}}^{\quad
r})^{-1/2}$. Using the Cauchy-Schwarz inequality
\[
\left| \mbox{Tr}S_{w^{j}}{A_{j}}\right| ^{2}\leq \mbox{Tr}S_{w^{j}}{A_{j}}
^{\ast }{A_{j}},
\]
we obtain
\begin{equation}
P_{j}({\cal W},{\bf X})\leq 2(1-\mbox{Tr}S_{w^{j}}A_{j})=2[1-\mbox{Tr}
S_{w^{j}}^{1+r/2}(\sum_{l=1}^{M}S_{w^{l}}^{\quad r})^{-1/2}].  \label{e1}
\end{equation}

Let $S_{w^{j}}=\sum_{\alpha }\lambda _{j}^{\alpha }|e_{j}^{\alpha }\rangle
\langle e_{j}^{\alpha }|$ be the spectral decomposition of the operator $
S_{w^{j}}$, then (\ref{e1}) takes the form
\begin{equation}
P_{j}({\cal W},{\bf X})\leq 2\sum_{\alpha }\lambda _{j}^{\alpha }\langle
e_{j}^{\alpha }|\left[ I-\left( \frac{\sum_{l=1}^{M}S_{w^{l}}^{\quad r}}{
(\lambda _{j}^{\alpha })^{r}}\right) ^{-1/2}\right] |e_{j}^{\alpha }\rangle .
\label{e2}
\end{equation}
Applying the inequality
\begin{equation}
2(1-x^{-1/2})\leq (x-1),\quad x>0,  \label{in}
\end{equation}
we obtain
\begin{equation}
2\left[ I-\left( \frac{\sum_{l=1}^{M}S_{w^{l}}^{\quad r}}{(\lambda
_{j}^{\alpha })^{r}}\right) ^{-1/2}\right] \leq \left( \frac{
\sum_{l=1}^{M}S_{w^{l}}^{\quad r}}{(\lambda _{j}^{\alpha })^{r}}\right) -I
\end{equation}
\begin{equation}
=(\lambda _{j}^{\alpha })^{-r}\sum_{l\neq j}S_{w^{l}}^{\quad r}+\sum_{\beta
\neq \alpha }\left[ \left( \frac{\lambda _{j}^{\beta }}{\lambda _{j}^{\alpha
}}\right) ^{r}-1\right] |e_{j}^{\beta }\rangle \langle e_{j}^{\beta }|
\label{h2}
\end{equation}
By substituting this into (\ref{e2}), we see that for $0<r\leq 1$
\begin{equation}
P_{j}({\cal W},{\bf X})\leq \mbox{Tr}S_{w^{j}}^{1-r}\sum_{l\not{=}
j}S_{w^{l}}^{\quad r},  \label{e4}
\end{equation}
in particular, for $r=1/2$ we obtain (\ref{e5}). By continuity argument we
can drop the assumption of nondegeneracy of the operators $S_{w^{j}}.\Box $

{\sl Corollary:} For any collection of states $S_{i};i=1,\dots a$
\begin{equation}
C \geq -\log \min_{\pi }\mbox{Tr}\left[ \sum_{i=1}^{a}\pi _{i}\sqrt{S_{i}}
\right] ^{2}.  \label{h1}
\end{equation}

{\sl Proof.} Let us apply random coding. Then from (\ref{e5}) using the fact
that the words are i.i.d., we find
\begin{equation}
{\sf E}\min_{{\bf X}}{\bar{P}}({\cal W},{\bf X})\leq (M-1)\mbox{Tr}({\sf E}
\sqrt{S_{w^{j}}})^{2}.  \label{h6}
\end{equation}
The expectation is
\[
\mbox{Tr}({\sf E}\sqrt{S_{w^{j}}})^{2}=\mbox{Tr}\left[ \left( \sum_{i}\pi
_{i}\sqrt{S_{i}}\right) ^{\otimes n}\right] ^{2}=\left[ \mbox{Tr}\left(
\sum_{i}\pi _{i}\sqrt{S_{i}}\right) ^{2}\right] ^{n},
\]
which gives (\ref{f27}) with $s=1,c=1.$ Choosing $M=2^{nR}$, we get \ (\ref
{h1}). $\Box $

{\sl Remark:} The above proof did not involve the quantum coding theorem
(\ref{cap}). On the other hand, by letting $s=1$ in (\ref{cap1}) we obtain
inequality
\begin{equation}
H\left( \sum_{i=1}^{a}\pi _{i}{S}_{i}\right) -\sum_{i=1}^{a}\pi _{i}H({S}
_{i})\geq -\log \mbox{Tr}\left( \sum_{i=1}^{a}\pi _{i}\sqrt{S_{i}}\right)
^{2}  \label{ineq}
\end{equation}
which, combined with (\ref{cap}), also gives (\ref{h1}).

The quantity in the right-hand side of (\ref{h1}) is a quantum analog of the
{\sl cutoff rate} widely used in applications of information theory (see
\cite{hirota}). Since it is easier to calculate than the capacity (the
minimum over $\pi $ can be evaluated explicitly), it can be used as a
practical lower bound. In particular, consider a quantum-quantum channel $
\Phi $, which is a completely positive trace preserving map of states, and
substitute $S_{i}=\Phi ^{\otimes n}(T_{i})$, where $T_{i}$ are the input
states in ${\cal H}^{\otimes n}$ to be optimized after. Then (\ref{h1})
implies a lower bound for the classical capacity of the channel, which might
be relevant to the problem of additivity of the capacity with respect to
entangled inputs \cite{ben}, \cite{hol73}, although the problem of
additivity of the cutoff rate is itself by no means simple.

Finally let us show how the bound (\ref{f27}) can be obtained for {\sl
commuting} $S_{i}$ along these lines. Taking expectation of (\ref{e2}), we
get
\begin{equation}
{\sf E}P_{j}({\cal W},{\bf X})\leq {\sf E}\sum_{\alpha }\lambda _{j}^{\alpha
}\langle e_{j}^{\alpha }|2{\sf E}\left\{ \left. \left[ I-\left( \frac{
\sum_{l=1}^{M}S_{w^{l}}^{\quad r}}{(\lambda _{j}^{\alpha })^{r}}\right)
^{-1/2}\right] \right| w^{j}\right\} |e_{j}^{\alpha }\rangle ,  \label{g2}
\end{equation}
where inside is the conditional expectation with respect to the fixed word $
w^{j}$. Taking this conditional expectation in (\ref{h2}), we get
\begin{equation}
{\sf E}\left\{ \left. 2\left[ I-\left( \frac{\sum_{l=1}^{M}S_{w^{l}}^{\quad
r}}{(\lambda _{j}^{\alpha })^{r}}\right) ^{-1/2}\right] \right|
w^{j}\right\}
\end{equation}
\begin{equation}
\leq (\lambda _{j}^{\alpha })^{-r}(M-1){\sf E}S_{w^{l}}^{\quad
r}+\sum_{\beta \neq \alpha }\left[ \left( \frac{\lambda _{j}^{\beta }}{
\lambda _{j}^{\alpha }}\right) ^{r}-1\right] |e_{j}^{\beta }\rangle \langle
e_{j}^{\beta }|.
\end{equation}
On the other hand, the left hand side is less or equal than $2I$. If all
operators commute, all the matrices are diagonal in
the basis $\left\{ e_{j}^{\beta }\right\} $ which is the same
for all $j ,$ and we can use inequalities $\min (2,x)\leq 2x^{s}$ and $
(x+y)^{s}\leq x^{s}+y^{s}$, valid for $x,y\geq 0$ and $0\leq s\leq 1$, to
obtain that the left hand side does not exceed
\[
2\left[ (\lambda _{j}^{\alpha })^{-rs}(M-1)^{s}\left( {\sf E}
S_{w^{l}}^{\quad r}\right) ^{s}+\sum_{\beta \neq \alpha }\left[ \left( \frac{
\lambda _{j}^{\beta }}{\lambda _{j}^{\alpha }}\right) ^{r}-1\right]
^{s}|e_{j}^{\beta }\rangle \langle e_{j}^{\beta }|\right] .
\]
Substituting this into (\ref{g2}), we obtain
\begin{equation}
{\sf E}P_{j}({\cal W},{\bf X})\leq {\sf E}\sum_{\alpha }(\lambda
_{j}^{\alpha })^{1-rs}\langle e_{j}^{\alpha }|2(M-1)^{s}\left( {\sf E}
S_{w^{l}}^{\quad r}\right) ^{s}|e_{j}^{\alpha }\rangle
\end{equation}
\begin{equation}
=2(M-1)^{s}\mbox{Tr}\left( {\sf E}S_{w^{j}}^{\quad 1-rs}\right) \left( {\sf E
}S_{w^{l}}^{\quad r}\right) ^{s}.
\end{equation}
Choosing $r=\frac{1}{1+s}$ this gives
\begin{equation}
{\sf E}P_{j}({\cal W},{\bf X})\leq 2(M-1)^{s}\mbox{Tr}\left( {\sf E}
S_{w^{j}}^{\quad \frac{1}{1+s}}\right) ^{1+s},
\end{equation}
whence (\ref{f27}) follows.

Moreover, we can omit the factor 2, if we use commutativity from the start,
avoid the Cauchy-Schwarz inequality and use $1-x^{-1}\leq x-1,x>0,$ instead
of (\ref{in}).

\begin{center}
\centerline{\sc III. The expurgation lower bound}
\end{center}

As it is well known in the classical information theory, for low rates $R$
codes with high probability of error become to dominate in the random coding
ensemble. In order to reduce the influence of choosing such bad codes
ingenious {\sl expurgation} technique has been developed, see \cite{gal},
Ch. 5.7.

{\sl Theorem:} For arbitrary density operators $S_{i}$ the expurgation bound
holds:
\begin{equation}
\min_{{\cal W},{\bf X}}P_{max}({\cal W},{\bf X})\leq \inf_{s\geq 1}\left(
4(M-1)\left[ \sum_{i,k=1}^{a}\pi _{i}\pi _{k}(\mbox{Tr}\sqrt{S_{i}}\sqrt{
S_{k}})^{\frac{1}{s}}\right] ^{n}\right) ^{s}.  \label{f28}
\end{equation}

{\sl Proof}. Using (\ref{e5}) and the inequality $(\sum a_{i})^{r}\leq
\sum a_{i}^{r},0<r\leq 1,$ we obtain for $s\geq 1$

\begin{equation}
(P_{j}({\cal W},{\bf X}))^{1/s}\leq (\sum_{l\not=j}\mbox{Tr}\sqrt{S_{w^{j}}}
\sqrt{S_{w^{l}}})^{1/s}\leq \sum_{l\not=j}(\mbox{Tr}\sqrt{S_{w^{j}}}\sqrt{
S_{w^{l}}})^{1/s}.  \label{e9}
\end{equation}
We again apply the Shannon$^{\prime }$s random coding scheme, assuming that
the codewords are chosen at random, independently and with the probability
distribution (\ref{iii5}) for each word. We start with an ensemble of codes
with $M^{\prime }=2M-1$ codewords. Then according to the Lemma from Ch. 5.7
\cite{gal} (which is a simple corollary of the central limit theorem) there
exists a code in the ensemble of codes with $M^{\prime }=2M-1$ codewords,
for which at least $M$ codewords satisfy
\begin{equation}
P_{j}({\cal W},{\bf X})\leq \left[ 2{\sf E}P_{j}({\cal W},{\bf X})^{1/s}
\right] ^{s},  \label{e8}
\end{equation}
for arbitrary $s\geq 1$ (without loss of generality we can assume that (\ref
{e8}) holds for $j=1,\ldots ,M$). Then taking into account that $M^{\prime
}-1=2(M-1)$, we have from (\ref{e9})
\begin{equation}
P_{j}({\cal W},{\bf X})\leq \left[ 4(M-1){\sf E}(\mbox{Tr}\sqrt{S_{w^{j}}}
\sqrt{S_{w^{l}}})^{1/s}\right] ^{s}.  \label{e7}
\end{equation}
Using the fact that the words are i.i.d., we find
\[
{\sf E}(\mbox{Tr}\sqrt{S_{w^{j}}}\sqrt{S_{w^{l}}})^{1/s}
\]
\[
=\sum_{i_{1},...i_{n};j_{1},...,j_{n}}\pi _{i_{1}}\dots \pi _{i_{n}}\pi
_{j_{1}}\dots \pi _{j_{n}}(\mbox{Tr}\sqrt{S_{i_{1}}}\sqrt{S_{j_{1}}}
)^{1/s}\dots (\mbox{Tr}\sqrt{S_{i_{n}}}\sqrt{S_{j_{n}}})^{1/s}
\]
\begin{equation}
=\left[ \sum_{i,j}\pi _{i}\pi _{j}(\mbox{Tr}\sqrt{S_{i}}\sqrt{S_{j}})^{1/s}
\right] ^{n},
\end{equation}
whence the theorem follows.$\Box $

Again, it is convenient to introduce the function
\[
{\tilde{\mu}}(\pi ,s)=-s\log \sum_{i,k=1}^{a}\pi _{i}\pi _{k}(\mbox{Tr}
\sqrt{S_{i}}\sqrt{S_{k}})^{\frac{1}{s}},
\]
then taking $M=2^{nR}$, we obtain the expurgation lower bound for the
reliability function
\[
E(R)\geq \max_{\pi }\sup_{s\geq 1}({\tilde{\mu}}(\pi ,s)-sR)\equiv
E_{ex}(R),
\]

The function ${\tilde{\mu}}(\pi ,s)$ is concave (see Appendix), increasing
from the value
\[
{\tilde{\mu}}(\pi ,1)=\mu (\pi ,1)=-\log \mbox{Tr}\left( \sum_{i=1}^{a}\pi
_{i}\sqrt{S_{i}}\right) ^{2}
\]
for $s=1$ to
\[
{\tilde{\mu}}(\pi ,\infty )=-\sum_{i,k=1}^{a}\pi _{i}\pi _{k}\log \mbox{Tr}
\sqrt{S_{i}}\sqrt{S_{k}},
\]
(which may be infinite).

By introducing
\begin{equation}
E_{ex}(\pi ,R)=\sup_{s\geq 1}[{\tilde{\mu}}(\pi ,s)-sR],  \label{f24a}
\end{equation}
we can investigate the behavior of $E_{ex}(\pi ,R)$ like in the classical
case. Namely, for $0<R\leq {\tilde{\mu}}^{\prime }(\pi ,1)$, where
${\tilde{\mu}}^{\prime }(\pi ,1)\leq {\tilde{\mu}}(\pi ,1)$
(see Appendix), the function $E_{ex}(\pi ,R)$ is concave, decreasing from
\begin{equation}
E_{ex}(\pi ,+0)={\tilde{\mu}}(\pi ,\infty )  \label{f250}
\end{equation}
to $E_{ex}({\tilde{\mu}}^{\prime }(\pi ,1))={\tilde{\mu}}(\pi ,1)-{\tilde{\mu
}}^{\prime }(\pi ,1)$. In the interval ${\tilde{\mu}}^{\prime }(\pi ,1)\leq
R\leq {\tilde{\mu}}(\pi ,1)$ it is linear function
\[
E_{ex}(\pi ,R)={\tilde{\mu}}(\pi ,1)-R,
\]
and $E_{ex}(\pi ,R)=0$ for ${\tilde{\mu}}(\pi ,1)\leq R<C$.

Finally, let us evaluate the limiting value $E(+0)$ of the reliability
function at zero rate. We remind notation $|A|=\sqrt{A^* A}$ where $A$
is an operator in $\cal{H}$.

{\sl Proposition 2:} If $S_{i} S_{k}{\not =} 0 $
{\sl \ for any }$1\leq i,k\leq a$ then
\[
-\min_{\pi }\sum_{i,k=1}^{a}\pi _{i}\pi _{k}\log \mbox{Tr}\sqrt{S_{i}}
\sqrt{S_{k}}\leq E(+0)
\]
\begin{equation}
\leq -2\min_{\pi }\sum_{i,k=1}^{a}\pi _{i}\pi _{k}\log \mbox{Tr}\left| \sqrt{
S_{i}}\sqrt{S_{k}}\right| ,  \label{f26}
\end{equation}
If $S_{i} S_{k} =0$ for some $i,k,$ then $
E(+0)=\infty.$

{\sl Proof}. The proof is a generalization of that given in \cite{bur} for
the case of pure states. Note that in this case the left and right hand
sides of (\ref{f26}) coincide, giving the exact values of $E(+0)$.

From (\ref{f250}) we see that $E(+0)$ is greater than or equal to the
left hand side of (\ref{f26}). On the other hand,
\[
P_{max}({\cal W},{\bf X})\geq \max_{w\neq w^{\prime }}\min_{{\bf X}
}P(\left\{ S_{w},S_{w^{\prime }}\right\} ,{\bf X}),
\]
where $w,w^{\prime }$ are arbitrary two codewords from ${\cal W}$. The
minimal error probability of discrimination between the two equiprobable
states $S_{w},S_{w^{\prime }}$ is
\[
\min_{{\bf X}}P(\left\{ S_{w},S_{w^{\prime }}\right\} ,{\bf X})=\frac{1}{2}[
1-\frac{1}{2}\mbox{Tr}\left| S_{w}-S_{w^{\prime }}\right| ].
\]
(cf. \cite{hol}). In \cite{hol72} the following estimates were established
for the trace norm of the difference $S_{1}-S_{2}$ of any two density
operators $S_{1},S_{2}:$
\[
2(1-\mbox{Tr}\sqrt{S_{1}}\sqrt{S_{2}})\leq \mbox{Tr}\left|
S_{1}-S_{2}\right| \leq 2\sqrt{1-(\mbox{Tr}\sqrt{S_{1}}\sqrt{S_{2}})^{2}}.
\]
The proof of the second inequality can be easily modified to obtain
\[
\mbox{Tr}\left| S_{1}-S_{2}\right| \leq 2\sqrt{1-(\mbox{Tr}\left| \sqrt{S_{1}
}\sqrt{S_{2}}\right| )^{2}}
\] (see also \cite{fuchs}).
Therefore we get
\[
\min_{{\bf X}}P(\left\{ S_{w},S_{w^{\prime }}\right\} ,{\bf X})
\]
\[
\geq \frac{1}{2}\left[ 1-\sqrt{1-(\mbox{Tr}\left| \sqrt{S_{w}}\sqrt{
S_{w^{\prime }}}\right| )^{2}}\right] \geq \frac{1}{4}(\mbox{Tr}\left| \sqrt{
S_{w}}\sqrt{S_{w^{\prime }}}\right| )^{2},
\]
and
\[
P_{max}({\cal W},{\bf X})\geq \max_{w\neq w^{\prime }}\frac{1}{4}(\mbox{Tr}
\left| \sqrt{S_{w}}\sqrt{S_{w^{\prime }}}\right| )^{2}.
\]
It follows that
\[
E(+0)\leq -\lim_{n\rightarrow \infty }{\frac{2}{n}}\max_{w\neq w^{\prime
}}\log \mbox{Tr}\left| \sqrt{S_{w}}\sqrt{S_{w^{\prime }}}\right| .
\]
Repeating argument from the proof of Proposition 3 from \cite{bur}, we
obtain the second inequality in (\ref{f26}). $\Box $
\newpage
\centerline{\sc Acknowledgments}
\vskip10pt
The main part of this work was done
when the author was visiting the Institute for Mathematical Physics of the
Technical University of Braunschweig with the A. von Humboldt Research
Award. The author acknowledges the hospitality and stimulating discussions
with Prof. R. F. Werner. He is grateful to the referees for suggesting
several corrections to the initial version of the paper.
\vskip20pt
\centerline{\sc Appendix}
\vskip10pt
{\sl Properties of the functions }$\tilde{\mu}(\pi ,s), \mu (\pi ,s).$

1. Let us calculate derivatives of the function $\tilde{\mu}
(\pi ,s)$ with respect to $s$ and show that the first derivative is
nonnegative, while the second is nonpositive.  Denoting $F_{ik}(s)=(\mbox{Tr}\sqrt{S_{i}}\sqrt{
S_{k}})^{\frac{1}{s}},F(s)=\sum_{i,k=1}^{a}\pi _{i}\pi _{k}F_{ik}(s),$ we
have (with $\log $ denoting in what follows natural logarithms)
\[
\tilde{\mu}^{\prime}(\pi ,s)=-\log F(s)- sF(s)^{-1}F^{\prime}(s)
\]
\[
= F(s)^{-1}\sum_{i,k=1}^{a}\pi _{i}\pi _{k}F_{ik}(s)(\log F _{ik}(s)-\log
F(s)).
\]
Using the inequality
\begin{equation}
x(\log x-\log y)\geq x-y,\quad x,y>0  \label{log}
\end{equation}
(cf. Proposition 3.16 of \cite{petz} ), we see that indeed $\tilde{\mu}
^{\prime }(\pi ,s)\geq 0.$ Taking into account that $F^{\prime }(1)\leq 0,$
we also obtain $\tilde{\mu}^{\prime }(\pi ,1)\leq \tilde{\mu}(\pi ,1).$

The second derivative
\[
\tilde{\mu}^{\prime \prime }(\pi, s)=(s F(s)^2)^{-1}
\left[\left(\sum_{i,k=1}^{a}\pi _{i}\pi _{k}F_{ik}(s)\log F
_{ik}(s)\right)^2\right.
\]
\[
\left.- \sum_{i,k=1}^{a}\pi _{i}\pi _{k}F_{ik}(s)(\log F
_{ik}(s))^2\sum_{i,k=1}^{a}\pi _{i}\pi _{k}F_{ik}(s)\right]
\]
is nonpositive by Cauchy-Schwarz inequality.

2. Let us show that $\mu'(\pi, s)\geq 0$. Introducing operator valued function
\[
A(s)=\sum_{i=1}^{a}\pi _{i}S_{i}^{\frac{1}{1+s}},
\]
and letting $G(s)=\mbox{Tr}A(s)^{1+s},$ we have $\mu (\pi ,s)=-\log G(s)$,
so that
\[
\mu ^{\prime }(\pi ,s)=-G(s)^{-1}G^{\prime }(s).
\]
To calculate $G^{\prime }(s)$ we use a generalization of formula (3.17) from
\cite{petz}, namely
\begin{equation}
\frac{d}{ds}\mbox{Tr}f(s,A(s))=\mbox{Tr}f_{s}^{\prime }(s,A(s))+\mbox{Tr}
f_{A}^{\prime }(s,A(s))A^{\prime }(s).  \label{der}
\end{equation}
We then obtain
\begin{equation}
\frac{d}{ds}G(s)=-\mbox{Tr}A(s)^{s}\sum_{i=1}^{a}\pi _{i}S_{i}^{\frac{1}{1+s}
}\left[ \log S_{i}^{\frac{1}{1+s}}-\log A(s)\right] .  \label{deriv}
\end{equation}
From (\ref{log}) we have
\begin{equation}
y^{s}x(\log x-\log y)\geq y^{s}(x-y),\quad x,y>0,
\end{equation}
therefore by Proposition 3.16 from \cite{petz}
\[
-\frac{d}{ds}G(s)\geq \mbox{Tr}A(s)^{s}\sum_{i=1}^{a}\pi _{i}\left[ S_{i}^{
\frac{1}{1+s}}-A(s)\right] =0.
\]

In the classical case the function $\mu (\pi, s)$ is concave in $s$
\cite{gal}, Appendix 5B.
We conjecture this property holds also in the quantum case, and we
postpone this problem to a separate investigation.

3. To compute $E_{r}(R)$ according to the definition (\ref{er1}) it is
expedient to perform maximization with respect to $\pi $ first. Maximizing $
\mu (\pi ,s)$ is equivalent to minimizing
\[
G(\pi ,s)=\mbox{Tr}\left( \sum_{i=1}^{a}\pi _{i}S_{i}^{\frac{1}{1+s}}\right)
^{1+s}.
\]
By Proposition 3.1 of \cite{petz} this function is convex in $\pi ,$ which
makes the general criterium of Theorem 4.4.1 from \cite{gal} applicable.

From this theorem it follows that probability distribution $\pi $ minimizes $
G(\pi ,s)$ if and only if there exists a constant $c$ such that
\[
\frac{\partial G(\pi ,s)}{\partial \pi _{j}}\geq c,\quad j=1,\dots ,a,
\]
with equality for those $j,$ for which $\pi _{j}>0.$ After some computation,
this amounts to
\begin{equation}
\mbox{Tr}S_{j}^{\frac{1}{1+s}}\left( \sum_{i=1}^{a}\pi _{i}S_{i}^{\frac{1}{
1+s}}\right) ^{s}\geq \mbox{Tr}\left( \sum_{i=1}^{a}\pi _{i}S_{i}^{\frac{1
}{1+s}}\right) ^{1+s},  \label{cri}
\end{equation}
with the corresponding equalities.

By using this necessary and sufficient condition one shows, as in Example 4
of Sec.5.6 \cite{gal}, that for two parallel channels 1 and 2
\[
\max_{\pi _{12}}\mu _{12}(\pi _{12},s)=\max_{\pi _{1}}\mu _{1}(\pi
_{1},s)+\max_{\pi _{2}}\mu _{2}(\pi _{2},s),
\]
implying a corresponding additivity property for $E_{r}(R).$ This gives an
answer to a question posed by R. Ahlswede. It is worthwhile to remind that
the additivity property does not hold in general for ${\tilde \mu}(\pi , s)$
even in the classical case \cite{gal}, Problem 5.26.

\end{document}